\documentclass{article}
\usepackage {graphicx}
\begin{document}
\baselineskip .75cm 
\begin{titlepage}
\title{\bf Self-consistent quasiparticle model\\ for 2, 3 and (2+1) flavor QGP}       
\author{Vishnu M. Bannur  \\
{\it Department of Physics}, \\  
{\it University of Calicut, Kerala-673 635, India.} }   
\maketitle
\begin{abstract}
Quasi-particle model of quark gluon plasma is the statistical mechanics of particles with medium dependent mass, related to plasma frequency, which was proposed to describe the thermodynamics of the medium itself. At relativistic limit the plasma frequency depends on  number density and temperature. The number density is a thermodynamic quantity of the medium which in turn depends on plasma frequency. Hence, one need to solve this problem self-consistently, instead of using perturbative expressions for plasma frequency. Here we carry out such a self-consistent calculations using our, recently developed, a new formulations of quasiparticle model. By adjusting a single parameter for each system, a remarkably good fit to lattice QCD results are obtained for 2, 3 and (2+1) flavor quark gluon plasma systems, first,  with zero chemical potential. Then, it is extended to systems with finite chemical potential and fits very well the lattice results without any new parameter.     
\end{abstract}
\vspace{1cm}
                                                                                
\end{titlepage}
\section{Introduction :}
Recently proposed self-consistent quasiparticle model for quark gluon plasma (qQGP) \cite{ba.1} is extended to 2-flavor (2-f), 3-flavor (3-f) and (2+1)-flavor ((2+1)-f) QGP and is found to explain very well the lattice gauge theory (LGT) QCD results \cite{la.1}. Our model differs from other qQGP models \cite{oq.1} in many aspects. Firstly, our formulation is based on the standard statistical mechanics (SM) \cite{ba.2}, without the need of any reformulation of SM \cite{go.1}. Here we start from energy density ($\varepsilon$) and  particle number density ($n$), well defined in grand canonical ensemble (GCE) and then all other thermodynamic (TD) quantities, including pressure, are derived from $\varepsilon$, $n$  and using the standard TD relations. Hence, there is no TD inconsistency in our model \cite{ba.3}. Secondly, the thermal mass of quasiparticle, in our model, is assumed to be function of density which in term depends on thermal mass and hence one need to solve it  self-consistently \cite{ba.1} with the requirements that at high temperature thermal masses go to that of QCD perturbative results. Note that all other qQGP models \cite{oq.1} used QCD perturbative thermal masses which is appropriate to ideal system, because one uses ideal thermal propagators in these calculations, but the system is non-ideal. Only at high temperature it may be approximated to ideal system. Thirdly, there is only single system dependent adjustable parameter and gives a surprisingly good fit to LGT results. We will discuss and compare 3 models, in our formulation, based on 3 forms of thermal mass. Model-I is based on QCD perturbative thermal mass without self-consistent calculation. Model-II is based on our self-consistent calculation, using the simpler, approximate dispersion relation ($\omega_k$ Vs $k$). In model-III, we repeat the self-consistent calculations of Model-II, but using the dispersion relation of hard-thermal-loop (HTL) resumed QCD perturbation theory. We will see that the results of Model-I, is not satisfactory, but results improve considerably as we go from Model-I to Model-III, because of our self-consistent calculations. 

\section{New formulation of qQGP:} 

As in any qQGP model, here also, we assume that the thermal properties of interacting real particles are modeled by non-interacting quasiparticles with additional thermal mass which depends on thermodynamic quantities due to collective properties of the medium. Following the the standard SM, in GCE, $\varepsilon$ and $n$ are the only TD quantities, defined as the  average of the energy and the number of particles. All other TD quantities, including pressure, need to be derived \cite{pa.1,ba.2}. In usual SM, the masses of particles are zero or constant and 
hence one can derive the expression for the pressure from $\varepsilon$, using TD relations,  and find that the pressure is equal to the logarithm of the partition function \cite{pa.1}. But, here in QGP it is not true, since the thermal mass depends on TD quantities which in turn depends on thermal mass through a functional relationship and hence whole problem need to be solved self-consistently. 

The energy density is \cite{ba.2} 
\begin{equation}
\varepsilon = \frac{g_f}{2 \pi^2} \int_0^{\infty} dk\,k^2 \frac{\omega_k}{(z^{-1} e^{\frac{\omega_k}{T}} \mp 1)} \;\;,
\end{equation}
where $g_f$ is the degeneracy and $\mp$ refers to bosons and fermions. $z$ is the fugacity.  
Note that, here we have assumed that the whole thermal energy is used to excite quasiparticles and hence the vacuum energy contribution is neglected. A detailed discussion on the vacuum energy contribution and TD consistency relation was commented in Ref. \cite{ba.3}. 
Similarly, the number density is 
\begin{equation}
n = \frac{g_f}{2 \pi^2} \int_0^{\infty} dk\,k^2 \frac{1}{(z^{-1} e^{\frac{\omega_k}{T}} \mp 1)} \;\;.
\end{equation}
Single particle energy or $\omega_k$ depends on thermal masses and momentum $k$. The exact expression for $\omega_k$ may be obtained by solving the approximate dispersion relation,  obtained using HTL calculations \cite{bk.1}, 
\begin{equation}
\omega_k^2 = k^2 + \frac{3}{2} \, \omega_p^2 \,\left( \frac{\omega^2}{k^2} + 
(1 - \frac{\omega^2}{k^2}) \,\frac{\omega}{2 \, k} \,
\ln \left|\frac{\omega+k}{\omega - k} \right| \right) \,\, .  \label{eq:drg}
\end{equation} 
for gluons and 
\begin{equation}
\omega_k = k + \frac{m_f^2}{k}\,\left( 1 - \frac{\omega_k - k}{2 k} \, \ln \left|\frac{\omega_k + k}{\omega_k - k} \right| \right) \,\, .  \label{eq:drq} 
\end{equation}    
for quarks, where $\omega_p$ and $m_f$ are plasma frequencies related to thermal masses. Both of them are, in general, functions of temperature ($T$) and chemical potential ($\mu$). First, we consider QGP system with $\mu = 0$ and at the end with $\mu \ne 0$.    
Above dispersion relation may be further approximated to a simpler form,
\begin{equation}
\omega_k = \sqrt{k^2 + m_g^2} \;\;, \label{eq:dr1}
\end{equation} 
and 
\begin{equation}
\omega_k = \sqrt{k^2 + m_q^2} \;\;, \label{eq:dr2} 
\end{equation} 
for gluons and quarks respectively, valid at high temperatures. The thermal masses are defined as 
\begin{equation} 
m_g^2 \equiv \frac{3}{2} \omega_p^2 \;\;\;\mbox{and}\;\;\; m_q^2 \equiv 2 m_f^2 \;\;, 
\end{equation}
respectively, for massless particles. For massive quark with mass $m_0$, $m_q^2$ may be modified as \cite{oq.1}  
\begin{equation}
m_q^2 = (m_0 + m_f)^2 + m_f^2 \;\;. 
\end{equation}  

\begin{figure}[h]
\centering
\includegraphics[height=8cm,width=12cm]{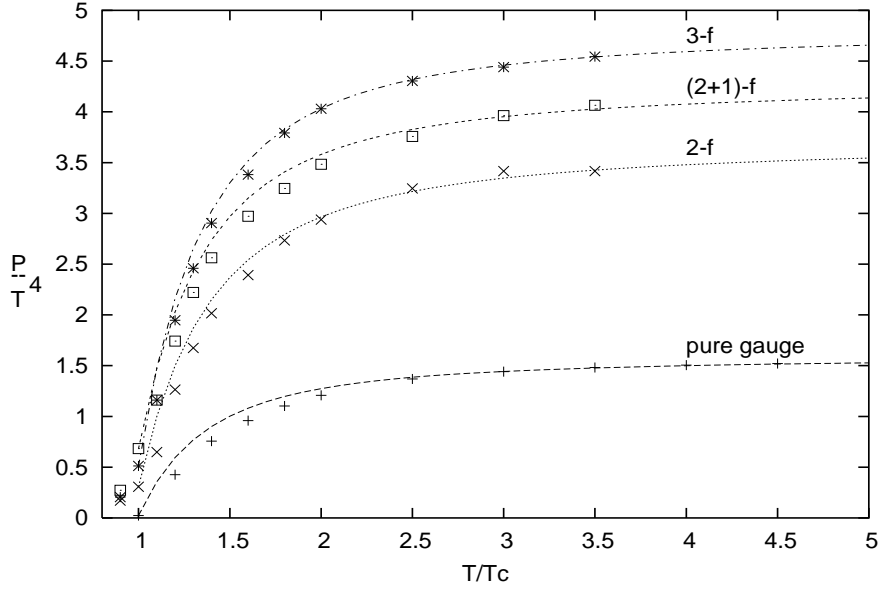}
\caption{ Plots of $P/ T^4 $ as a function of $T/T_c$ from the Model-I and lattice results \cite{la.1} (symbols) for pure gauge, 2-f, 3-f and (2+1)-f QGP.} 
\end{figure}

\begin{figure}[h]
\centering
\includegraphics[height=8cm,width=12cm]{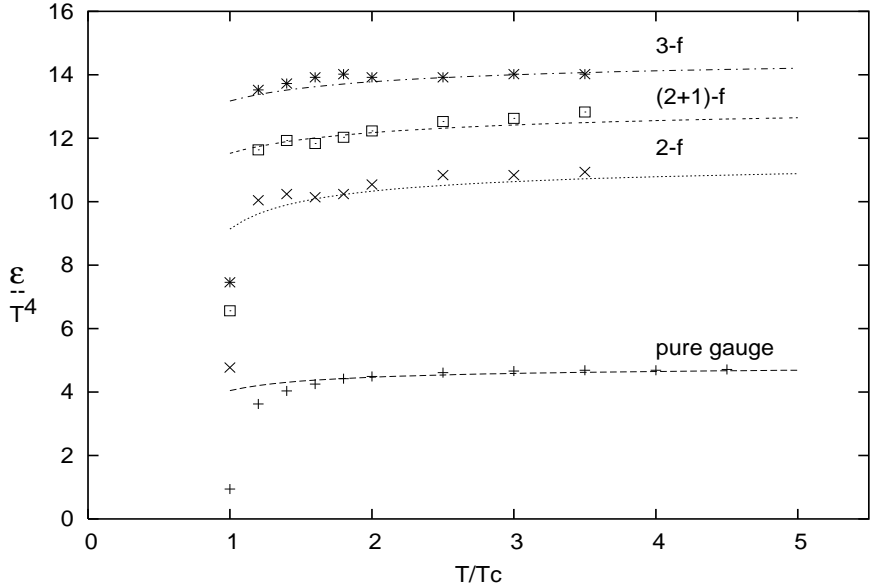}
\caption{ Plots of $\varepsilon/ T^4 $ as a function of $T/T_c$ from the Model-I and lattice results \cite{la.1} (symbols) for pure gauge, 2-f, 3-f and (2+1)-f QGP.} 
\end{figure}

First, let us consider our Model-I where, just like in other qQGP models, we take 
\begin{equation}
\omega_p^2 = \frac{g^2\,T^2}{18}\,(6 + n_f) \;\;, \label{eq:wp}
\end{equation}
and 
\begin{equation}
m_f^2 = \frac{g^2\,T^2}{6} \;\;, \label{eq:mf} 
\end{equation}
where $g$ is the QCD coupling constant and $n_f$ is the number of flavors.  These are perturbative results, valid for very high temperature, but are modified to take account of non-perturbative effects by using appropriate running coupling constant, $\alpha_s(T) \equiv \frac{g^2}{4 \pi}$. Different authors \cite{oq.1} use different phenomenological models of $\alpha_s$ and even thermal masses, with few adjustable parameters. Here, in our Model-I, we use 2-loop approximate running coupling constant, 
\begin{equation} 
\alpha_s (T) = \frac{6 \pi}{(33-2 n_f) \ln (T/\Lambda_T)}
\left( 1 - \frac{3 (153 - 19 n_f)}{(33 - 2 n_f)^2}
\frac{\ln (2 \ln (T/\Lambda_T))}{\ln (T/\Lambda_T)}
\right)  \label{eq:ls} \;, 
\end{equation}
where $\Lambda_T$ is the only one adjustable parameter in our model. Results are presented in Fig. 1 and Fig. 2. The general expression for the energy density is
\begin{equation}
\varepsilon = \frac{g_g}{2 \pi^2} \int_0^{\infty} dk\,k^2 \frac{\omega_k}{( e^{\frac{\omega_k}{T}} - 1)} + \frac{12\,n_u^{eff}}{2 \pi^2} \int_0^{\infty} dk\,k^2 \frac{\omega_k}{(e^{\frac{\omega_k}{T}} + 1)} +  \frac{12\,n_s^{eff}}{2 \pi^2} \int_0^{\infty} dk\,k^2 \frac{\omega_k}{( e^{\frac{\omega_k}{T}} + 1)} \;\;, \label{eq:eg} 
\end{equation}
where $g_g = 16$ is the degeneracy of gluons and $n_u^{eff}$, $n_s^{eff}$ are the effective number of flavors corresponding to light quarks ($u$ and $d$) and strange quark respectively. $n_f^{eff} = n_f$ for massless quarks and is less than $n_f$ for quarks with finite mass \cite{la.1}.

The pressure is obtained from the thermodynamic relation \cite{ba.1}
\begin{equation}
\varepsilon =  T \frac{\partial P}{\partial T} 
- P \,\, ,  \label{eq:td} 
\end{equation}
on integration with one integration constant which may be chosen such that $P \rightarrow P_s$, the Stefan-Boltzmann limit, as $T \rightarrow \infty$. However, it is difficult to adopt it in our numerical work and hence we fix it at $T=T_c$ to LGT results. 

\section{Self-consistent qQGP models - Model-II and Model-III:} 

To improve Model-I, in Model-II, we replace the perturbative expression for plasma frequencies, Eq. (\ref{eq:wp}) and Eq. (\ref{eq:mf}), by density dependent expression, 
\begin{equation}
\omega_p^2 = a_g^2\,g^2\,\frac{n_g}{T} + a_q^2\,g^2\,\frac{n_q}{T} \;\;, 
\end{equation}
for gluons and 
\begin{equation}
m_f^2 = b_q^2\,g^2\,\frac{n_q}{T} \;\;, 
\end{equation}
which is motivated from similar work in relativistic electron-positron plasma \cite{me.1,ba.4} where we know that the plasma frequency, at relativistic limit, is proportional to $n/T$. We fix the constant of proportionality $a_g$, $a_q$ and $b_q$ by demanding that as $T \rightarrow \infty$, $\omega_p$ and $m_f$ goes to the corresponding perturbative results. Since now thermal masses depends on density, the expression for density becomes a functional, given by
\begin{equation}
\frac{n}{T^3} = \frac{g_f}{2 \pi^2} \int_0^{\infty} dx\,x^2 \frac{1}{(z^{-1} e^{\omega} \mp 1)}
\end{equation}
where $x \equiv k/T$ and $\omega \equiv \omega_k /T$ which depends on $x$, $\frac{n_g}{T^3}$,  $\frac{n_q}{T^3}$ and $g^2 (T)$. 

\begin{figure}[h]
\centering
\includegraphics[height=8cm,width=12cm]{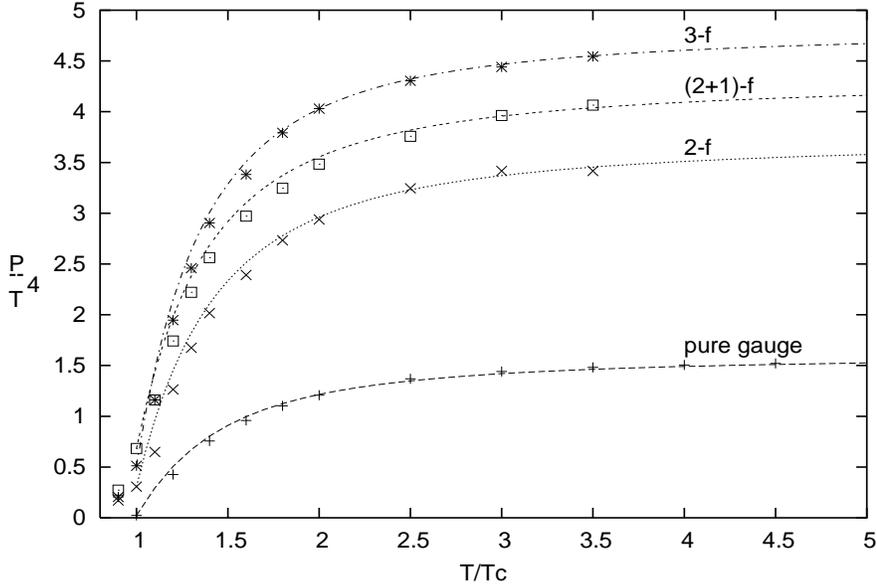}
\caption{ Plots of $P/ T^4 $ as a function of $T/T_c$ from the Model-II and lattice results \cite{la.1} (symbols) for pure gauge, 2-f, 3-f and (2+1)-f QGP.} 
\end{figure}

\begin{figure}[h]
\centering
\includegraphics[height=8cm,width=12cm]{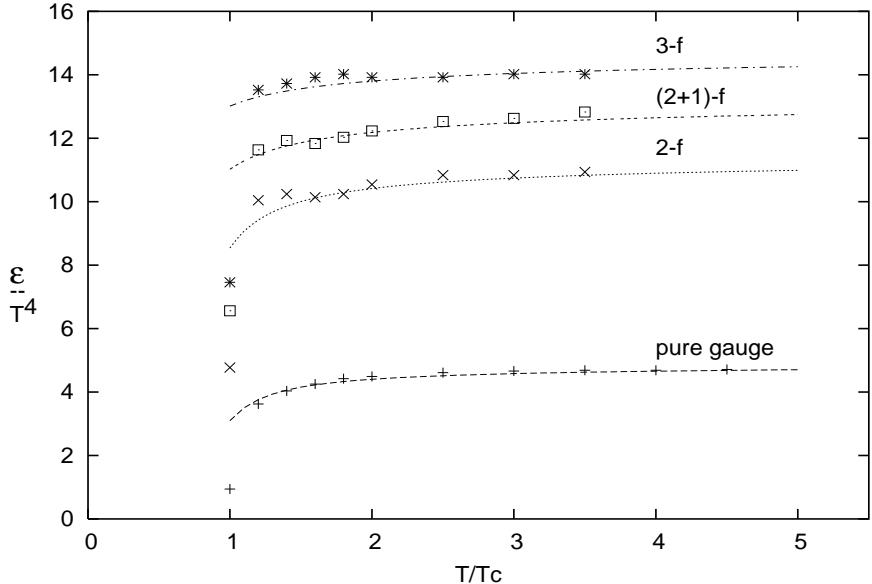}
\caption{ Plots of $\varepsilon/ T^4 $ as a function of $T/T_c$ from the Model-II and lattice results \cite{la.1} (symbols) for pure gauge, 2-f, 3-f and (2+1)-f QGP.} 
\end{figure}

So we need to solve density equation self-consistently to get the density and then we may calculate plasma frequencies which may be used to evaluate $\varepsilon$, $P$ and other TD quantities using Eq. (\ref{eq:eg}) and Eq. (\ref{eq:td}). 
It may be convenient to work with  
\begin{equation}
f_g^2 \equiv \int_0^{\infty} dx\,x^2 \frac{1}{( e^{\omega} - 1)} \;\;, 
\end{equation}
for gluons,  
\begin{equation}
f_q^2 \equiv \int_0^{\infty} dx\,x^2 \frac{1}{(z^{-1} e^{\omega} + 1)} \;\;, \label{eq:fq}
\end{equation}
for quarks and 
\begin{equation}
f_{\bar{q}}^2 \equiv \int_0^{\infty} dx\,x^2 \frac{1}{(z e^{\omega} + 1)} \;\;, \label{eq:fqb} 
\end{equation}
for antiquarks, 
where $f_g^2 \equiv \frac{n_g}{T^3} \; \frac{2 \pi^2}{g_g}$, $f_q^2 \equiv \frac{n_q}{T^3} \; \frac{2 \pi^2}{g_q}$ and $f_{\bar{q}}^2 \equiv \frac{n_{\bar{q}}}{T^3} \; \frac{2 \pi^2}{g_q}$. $g_q = 6 n_f^{eff}$ is the degeneracy of quarks and antiquarks.  It is easy to see that at high temperature $g^2 \rightarrow 0$ and hence $f_g^2 \rightarrow 2\,\zeta(3)$ and   $f_q^2 = f_{\bar{q}}^2  \rightarrow 2\,\eta(3)$ for $\mu = 0$ systems, appropriate to ideal plasma. This high temperature limit is taken to evaluate the coefficients $a_g$, $a_q$ and $b_q$ by comparing with perturbative results for plasma frequencies. 
Results are presented in Fig. 3 and Fig. 4. 

\begin{figure}[h]
\centering
\includegraphics[height=8cm,width=12cm]{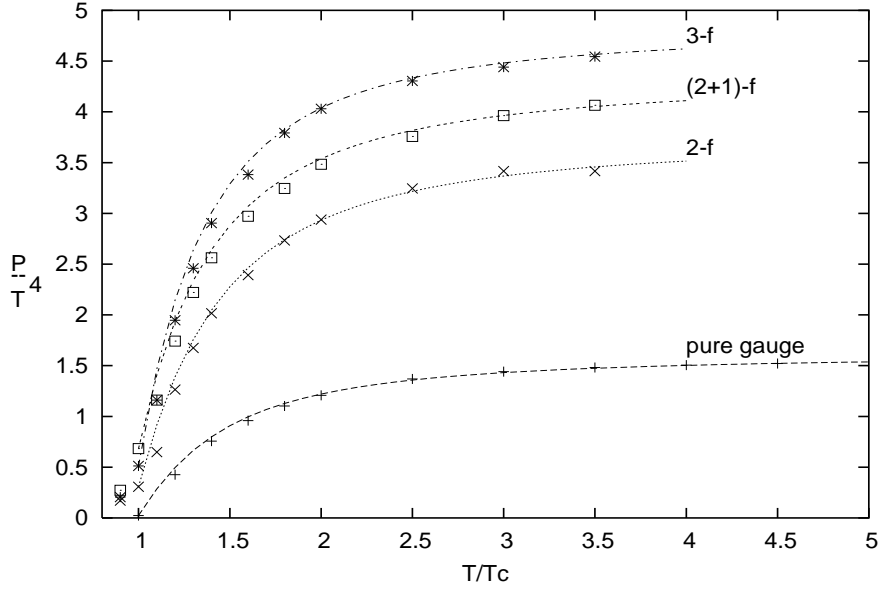}
\caption{ Plots of $P/ T^4 $ as a function of $T/T_c$ from the Model-III and lattice results \cite{la.1} (symbols) for pure gauge, 2-f, 3-f and (2+1)-f QGP.} 
\end{figure}

\begin{figure}[h]
\centering
\includegraphics[height=8cm,width=12cm]{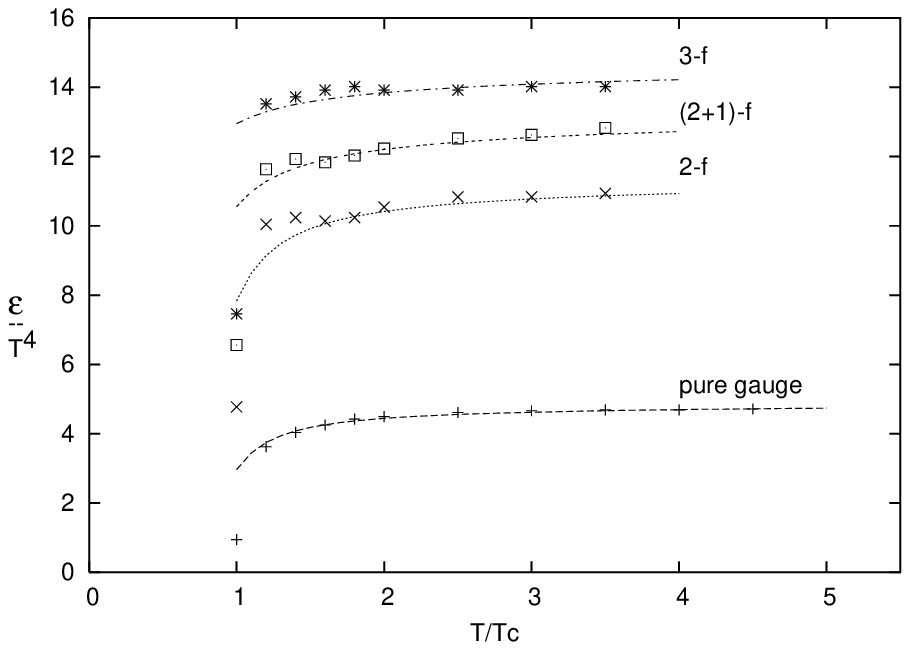}
\caption{ Plots of $\varepsilon/ T^4 $ as a function of $T/T_c$ from the Model-III and lattice results \cite{la.1} (symbols) for pure gauge, 2-f, 3-f and (2+1)-f QGP.} 
\end{figure}

Next we further improve Model-II by using the full HTL approximate dispersion relation, Eq. (\ref{eq:drg}) and Eq. (\ref{eq:drq}), instead of the approximate dispersion relation. 
First, one need to solve HTL dispersion relation for $\omega_k$ and then carry out the self-consistent calculations as done in Model-II. Results are presented in Fig. 5 and Fig. 6. 
We see that there is only a marginal improvement, especially near to $T < 2\,T_c$, as we go from Model-II to Model-III. Hence the dispersion relation of Model-II is a good approximation to that of HTL calculations.  

\section{Self-consistent qQGP with $\mu \ne 0$:} 

So far we have discussed QGP systems with $\mu = 0$ and found to explain LGT results very well using our self-consistent model. Let us apply our model to the system with $\mu \ne 0$. For $\mu \ne 0$, thermal masses and running coupling constant need to be modified as a function of $T$ and $\mu$. In our self-consistent model we demand that as $T \rightarrow \infty$, $m_f^2$ goes to QCD perturbative $(T, \mu)$ dependent functions, 
\begin{equation}
m_f^2 = \frac{g^2\,T^2}{6} (1 + \frac{\mu^2}{\pi^2 \,T^2} )\;\;.  
\end{equation}
The running coupling constant may be modified by replacing $T/\Lambda_T$ in Eq. (\ref{eq:ls})  by 
\begin{equation} 
\frac{T}{\Lambda_T} \,\sqrt{1 +  (1.91/2.91)^2 \,\frac{\mu^2}{T^2} } \,\,,  
\label{eq:lsa} 
\end{equation} 
using the results of Schneider \cite{sc.1}. Similar $\mu$ dependent modifications also used by Letessier, Rafelski \cite{ra.1}, but instead of $(1.91/2.91)^2$ factor they used $1/\pi^2$ which also gives similar results except near $T_c$. 
First, we solve the density equations, Eq. (\ref{eq:fq}) and Eq. (\ref{eq:fqb}), for quarks and antiquarks self-consistently and then subtract each other to get the net quark density $n_q /T^3$. Note that now the asymptotic values of $f_q$ and $f_{\bar{q}}$, needed to evaluate $m_f$, are not $2\,\eta(3)$, but given by 
\begin{equation}
f_q^2 = \sum_{l=1}^{\infty} (-1)^{l-1}\,\frac{z^l}{l^3} \;\;,  
\end{equation}
and 
\begin{equation}
f_{\bar{q}}^2 = \sum_{l=1}^{\infty} (-1)^{l-1}\,\frac{z^{-l}}{l^3} \;\;,    
\end{equation}
and both the expressions approach $2\,\eta (3)$ for $z = 1$. 
Results of our model, using Model-II, is plotted in Fig. 7 for $n_q/ T^3$ and fits LGT results without the need of any new parameters and $\Lambda_T$ is fixed earlier for $\mu =0$ system.  

\begin{figure}[h]
\centering
\includegraphics[height=8cm,width=12cm]{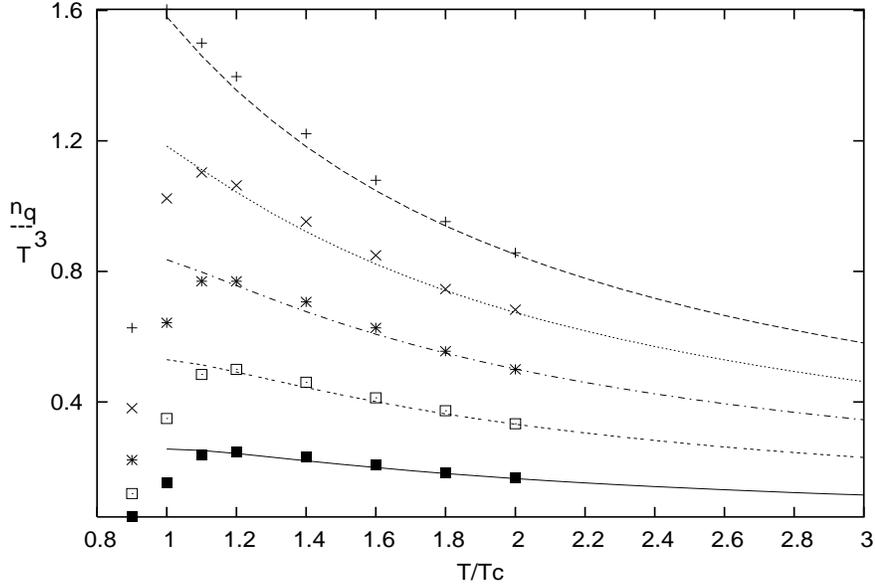}
\caption{ Plots of $n_q/ T^3 $ as a function of $T/T_c$ from the Model-II and lattice results \cite{la.1} (symbols) for 2-f QGP.} 
\end{figure}

\section{Results:} 

The analysis using Model-I is straight forward and the parameter $t_0 \equiv \frac{\Lambda_T}{T_c}$, is adjusted to get the best fit to LGT results on pressure. Using this $t_0$ we evaluate other TD quantities like $\varepsilon$. The fitted values of $t_0$ are 0.35, 0.4, 0.2, 0.2 for pure gauge, 2-f, 3-f and (2+1)-f QGP respectively and the results are plotted in Fig.1 and Fig. 2. In the case of massive quarks, effective number of quarks flavors, $n_f^{eff}$, is less than $n_f$ and may be obtained from the Ref. \cite{la.1} as $n_u^{eff} = 1.8344$ and $n_s^{eff} = 0.8275$. The LGT data \cite{la.1} for the systems with finite flavor were multiplied by a factor $1.1$ to take into account of the uncertainties in the continuum extrapolation. The fit to LGT results are not satisfactory for $T < 2.5\,T_c$, but fits well at higher temperature as it should. 
Note that the same model was also used by other qQGP studies \cite{oq.1} and obtained very good fit to LGT results, but adjusting 3 or more parameters, where as here we have only single adjustable parameter. In the case of Model-II, we solve the equation for density, self-consistently and use it to evaluate the pressure. Again, the parameter $t_0$ is adjusted to fit LGT results on pressure and then the energy density is evaluated. Now the fitted values of $t_0$ are higher and are 0.7, 0.6, 0.35, and 0.45 for pure gauge, 2-f, 3-f and (2+1)-f QGP respectively and the fit to LGT results, Fig. 3 and Fig. 4, improves a lot compared to Model-I. Similar self-consistent calculations is used in Model-III using the HTL dispersion relation, and the results are plotted in Fig. 5 and Fig. 6. The fitted values of $t_0$ are 0.74, 0.68, 0.4 and 0.55 for pure gauge, 2-f, 3-f and (2+1)-f QGP respectively. We see further improvement in the results, especially near $T < 2\,T_c$, and the values of $t_0$ are slightly higher. These values of $t_0$ may give better values of $\Lambda_T$, since Model-III is more accurate than Model-I and Model-II. In Fig. 7, results of our Model-II for 2-f QGP system at finite $\mu$ is compared with LGT results for quark density and reasonably good fit is obtained without any new parameter. Of course, very close to $T = T_c$ and for $T < T_c$, our quasiparticle model, which is based on plasma collective effects, may not explain the LGT results. One may need to include the effects of vacuum energy or effects of hadron resonance gas as done in Ref. \cite{bu.1}.   

\section{Conclusions:}

Here we have studied the thermodynamics of pure gauge, 2-f, 3-f and (2+1)-f QGP systems with our newly developed, self-consistent quasiparticle model using HTL approximate dispersion relation and it's approximate simpler dispersion relation.  By adjusting single system dependent adjustable parameter, $t_0 \equiv \Lambda_T /T_c$, we obtained a remarkable good fits to LGT results, without the need of too many adjustable parameters as required by other qQGP models \cite{oq.1}. For comparison, we have also studied these systems without using the self-consistent calculations, but fails to fit LGT results for $T < 2.5\, T_c$. From our calculation on $\mu \ne 0$ system, if LGT results are true, we need further more accurate dispersion relation than HTL approximate results to improve our model. It may be interesting to adopt this idea of self-consistent calculations in finite temperature QCD and solve the QCD at finite temperature without the need of qQGP.

\end{document}